# Slip band interactions and GND latent hardening in a galling resistant stainless steel

Benjamin Poole[a+], Fionn Dunne[a]

[a]Department of Materials, Imperial College London, Exhibition Road, London, SW7 2AZ UK

[+]Corresponding author, b.poole16@imperial.ac.uk

# Highlights

- Multiple slip systems are typically activated in Nitronic 60

- Increasing slip band interaction results in increased levels of lattice curvature and GND density

- Slip band blocking results in the highest generation of GND density, with slip band crossing resulting in an intermediate level of GND density. Planar slip bands produce low levels of GND density.

- GND accumulation adjacent to ferrite stringers and slip band interaction are the hardening mechanisms in Nitronic 60

- Crystal plasticity assessment confirms slip band blocking key to localised GND generation

# Abstract

Slip activation, slip band interactions, and GND densities in iron-base, galling resistant alloy Nitronic 60 have been characterised at the grain length scale using small-scale mechanical testing with high resolution digital image correlation and high-angular resolution electron backscatter diffraction. By correlating the two measurement techniques, new insight into slip band interactions, the generation of lattice curvature and the corresponding accumulation of geometrically necessary dislocations (GNDs) is provided. Multiple discrete slip bands are typically active within single grains, resulting in significant slip band interactions. Crossing slip bands were found to generate accumulations of GNDs. Regions where slip bands block other slip bands were associated with the highest GND densities, in excess of three time the densities of crossing slip bands. Representative crystal plasticity modelling investigations have demonstrated that discrete slip blocking events are responsible for locally elevated GND density. This behaviour is rationalised in terms of lattice curvature associated with the differing levels of constraint provided by the crossing or blocking-type behaviours. Ferrite grains are also found to contribute to the generation of GNDs. Together, these two effects provide significant work hardening mechanisms, likely to be key to the development of future iron-base hard facing alloys.

# 1 Introduction

Iron-base hard facing alloys are the main candidate to replace cobalt-base alloys found throughout the nuclear industry due to the radiological issues posed by cobalt-containing Stellite[TM] hard facings [1–3]. These alloys are required in a number of sliding wear applications and must withstand both wear and crucially galling, a severe surface-based plastic deformation mechanism which typically renders a component such as a valve inoperable [4]. Cobalt containing wear and corrosion products become entrained in the coolant flow within the primary circuits of light water reactors, becoming activated by the high neutron fluxes found in the reactor core. This gives rise to significant concentrations of radioactive $Co^{60}$, contributing to operator workplace radiation exposure. As such, the use of cobalt containing materials must be reduced so far as is reasonably practicable [5] whilst maintaining the high performance of cobalt-base hard facing alloys. A detailed understanding of the deformation mechanisms of iron-base hard facings is accordingly required.

In general, iron-base hard facing alloys cannot match the performance of their cobalt-base counterparts [2,3,6–9], particularly at elevated temperature [6,10–12]. A large body of literature regarding the performance of iron-base hard facings has developed over the last 60 years but much of it relies on large scale sliding wear tests, providing little insight into the deformation mechanisms at the microscale which will likely contribute to the overall galling resistance of these alloys. Modern hard facing alloys, such as RR2450 [13,14] or Nitromaxx [15], contain several different phases, resulting in complex deformation behaviour. A thorough understanding of the fundamentals of deformation in similar but more simple alloys is therefore required to truly understand the deformation of more complex systems.

Nitronic 60 is an ideal material to study the plastic deformation of iron-base galling resistant alloys. Nitronic 60 is primarily a galling resistant stainless steel, with a high nitrogen content to enhance its mechanical strength. The relatively simple microstructure, containing only austenite, a small volume of retained δ-ferrite and a trace amount of chromium carbide, allows for the deformation mechanisms within metallic phases to be studied. This microstructure facilitates the investigation of slip band interactions within grains. This is distinct to slip band-grain boundary interactions [16–19] with the formation of these structures driven by local plasticity through crystallographic slip, not the misorientation between neighbouring grains found at grain boundaries.

HR-DIC has been employed to study the deformation of Nitronic 60 and similar stainless steels. Zhao et al. [14] investigated Nitronic 60 alongside other iron-base alloys and cobalt-base Stellite 6.
Di Gioacchino and Fonseca [20,21] applied HR-DIC to small-scale tensile testing of 304L. These strain measurements were correlated with EBSD measured lattice rotations but HR-EBSD was not performed, and therefore there was no connection made with residual strain or GND density. Multiple different slip bands were found in some grains but the occurrence of slip bands impeding one another was not observed; in general slip bands were observed to intersect one another without any blocking. Polatidis et al. [22] examined 304 stainless steel under multi-axial loading and, again, found single slip to be prevalent with secondary and tertiary slip systems activating later on during deformation, at higher strain values. Again HR-EBSD was not performed so there was no linkage with lattice curvature or GND accumulation. HR-DIC has also been used to study 301 stainless steel with EBSD used to identify regions undergoing the γ → α′ strain induced martensite transformation but not to measure strain or GND density [23]. Whilst 300 series stainless steels display poor galling resistance, the formation of martensite is interesting as this is suggested to be beneficial in galling resistance [10,11,24,25]. In this study, the martensite transformation was observed at global strains in excess of 10%. Quantitative experimental evidence of strain induced martensitic phase transformations in Nitronic 60 could not be found in the literature, but it would be expected to behave in a similar way to the related iron-base hard facing alloy NOREM 02. NOREM 02 displays a characteristic disappearance in the γ → α′ strain induced transformation at temperatures above 180°C, which is often cited as the underlying cause of its reduction in elevated temperature galling resistance [10]. It would be expected that Nitronic 60 would also follow this trend with the formation of martensite more favourable at lower temperature.

Since Nitronic 60 is principally a specialist galling resistant alloy, it is relatively understudied compared to the more commonly used 300 series stainless steels. Nitronic 60 is itself a derivative of the 300 series stainless steels with the addition of manganese, silicon and crucially nitrogen. The nitrogen acts to improve the strength of Nitronic 60 but has the additional effect of reducing the stacking fault energy [26]; this may promote further strengthening mechanisms leading to enhanced galling resistance. A detailed understanding of the effect of nitrogen on the deformation of stainless steel is interesting scientifically and industrially, especially in the light of the EPRI developed high-nitrogen hard facing Nitromaxx [27] as this may provide a range of future iron-base hard facing alloys. It is therefore clear that an in-depth understanding of the plastic deformation of iron-base hard facing alloys is required for meaningful progress in their development to be made.

This paper presents a thorough investigation into the slip band interactions in Nitronic 60 using both HR-DIC and HR-EBSD at a high spatial resolution, providing insight at the sub-grain length scale. The different slip band interactions are characterised, and these varying behaviours are evaluated mechanistically with cross-correlation based GND density measurements. A representative crystal plasticity model is applied to gain further insight into these behaviours. Micromechanical mechanisms are proposed to explain how increasing intensity of slip band interaction leads to higher local levels of GND density and enhanced latent hardening.

## 2 Experimental methodology

### 2.1 Materials

Nitronic 60 is an iron-base galling resistant alloy with a high nitrogen concentration to enhance its mechanical properties and a high chromium concentration to provide corrosion resistance (composition in Table 1). It is designed to have a high work hardening rate to resist the gross plastic deformation found in highly loaded sliding contacts, particularly at elevated temperatures [28]. Corrosion resistance is important for nuclear power applications due to the corrosive primary coolant of light water reactors [1].

Nitronic 60 is mainly austenite with a small volume fraction of retained delta ferrite. The raw material was cast and extruded as a ~600 mm diameter rod and annealed at approximately 950°C by the supplier, resulting in the formation of ferrite stringers running parallel to the extrusion direction.

## 2.2 Specimen preparation and experimental procedure

A small scale three-point bend geometry (Figure 1a) was used for this study. The region of interest (ROI) is within the tensile fibre of the beam, generating an approximately horizontal tensile stress state.

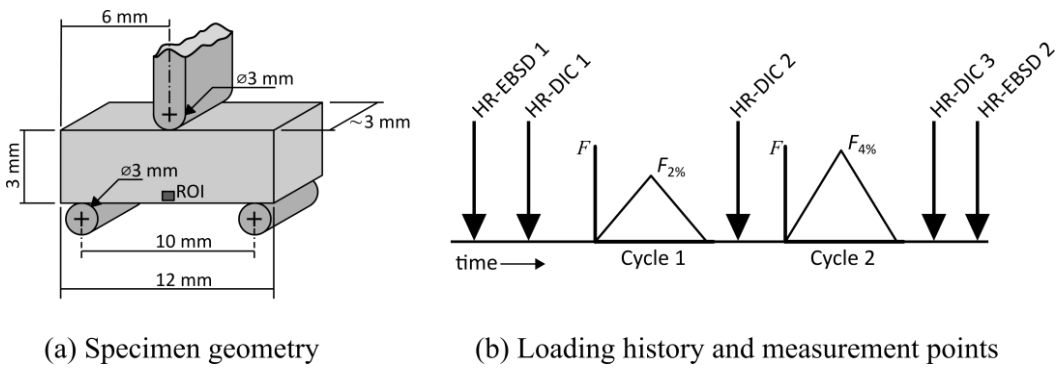

(a) Specimen geometry     (b) Loading history and measurement points

Figure 1: Loading conditions. (a) Geometry of the bend specimen with roller position and region of interest (ROI). Not to scale. (b) Schematic diagram of the experimental procedure. Pre- and post-deformation GND density and residual elastic strain measurements are captured during HR-EBSD 1 and HR-EBSD 2. Total strain maps using DIC are calculated by comparing images HR-DIC 2 with HR-DIC 1 to give measurements after cycle 1 and by comparing images HR-DIC 3 with HR-DIC 1 to give measurements after cycle 2. $F_{2\%}$ and $F_{4\%}$ denote the applied indentation force required to achieve 2% and 4% tensile strain respectively in the horiztonal direction region of interest.

The raw bar was electric discharge machined into $12 \times 3 \times 3.5$ mm specimens. The recast layer was removed by hand grinding to an 800 grit finish. The face of interest was ground to a 4000 grit finish with a series of silicon carbide grinding papers, followed by six minutes of fine grinding with 1 µm diamond suspension (Struers Nap-B1/MD-Nap pad). A final polishing step using a colloidal silica solution (30 minutes, 1:1 dilution of Struers OPS in deionised water) was applied to leave the front surface free of deformation. The final specimen thickness was approximately 3 mm. Microhardness indents were used as fiducial markers, allowing the region of interest to be quickly located and aligned.

Figure 1b details the experimental workflow, highlighting the stages required for HR-EBSD and HR-DIC. A Shimadzu AGS-X loading frame with a custom made three-point bend grip was used to apply two cycles of monotonic loading under force control. An isotropic elastic-plastic finite element model with the bulk material properties of Nitronic 60 was used to estimate the forces required to generate longitudinal strains of 2% and 4% for cycles 1 and 2 respectively. A HR-EBSD map was captured prior to all deformation over the region of interest. The gold remodelling technique [20,29] was used to apply a uniform layer of gold speckles over the region of interest and imaged prior to deformation and following cycles 1 and 2. Finally, the DIC speckles were removed and a second HR-EBSD scan was performed.

## 2.3  High-angular resolution electron backscatter diffraction

HR-EBSD maps were captured before and after testing (20 kV accelerating voltage, 120 μm aperture, 0.5 μm step size) over a region approximately 400 μm × 200 μm using a Zeiss Sigma 300 FESEM (field emission scanning electron microscope, Carl Zeiss AG, Germany) with a Bruker e⁻Flash EBSD camera (Bruker nano GmbH, Germany). HR-EBSD step sizes similar to this have been used to measure elastic strains in a range of materials [14,30–32]. It was not possible to capture a HR-EBSD map between the two cycles due to the presence of the gold speckles for HR-DIC. Patterns were recorded with a 2 × 2 binning giving a final pattern resolution of 800 × 600 pixels. Jiang et al. [33] have shown that a step size of 0.5 μm with a 4 × 4 detector binning to be acceptable for the measurement of GND densities. The step size, dwell time and size of the region scanned in this study represents a pragmatic balance between resolution and time taken to perform the measurements. The MTEX toolbox [34] was used to process EBSD data.

Residual elastic strains and GND densities were calculated using the cross-correlation technique based on the work of Wilkinson, Meaden and Dingley [35,36]. This technique applies image correlation techniques to compare individual electron backscatter diffraction patterns (EBSPs) to a reference pattern to calculate shifts. The pattern shifts are used to calculate the elastic deformation gradient tensor, from which elastic strains and GND densities can be calculated [37]. An in-house MATLAB code was used to perform this analysis; a full description is given by Jiang et al. [38].

Care must be taken interpreting the strain measurements acquired with this method. Since these measurements are taken by calculating EBSP shifts relative to a reference, all elastic strains are relative and not absolute values. Since EBSD measurements were taken in an unloaded state, the strains are also residual elastic strains. Therefore, strain values can only be compared *within* individual grains and no inferences can be made of the strain distributions *between* grains. GND density measurements are unaffected as they are based on lattice curvature measurements. Both step size and detector binning were kept consistent between scans [33].

## 2.4 High-resolution digital image correlation

The gold remodelling technique was used to provide features to track deformation [20,29]. A thin layer of gold was applied to the specimen surface following the initial EBSD scan using an Emitech-K575X gold sputter coating machine (Quorum Technologies ltd., UK). A deposition current of 20 mA and a time of 15 s was found to be suitable. The specimen was then placed in a moist environment, created with an upturned beaker on top of a smaller beaker containing deionised water on a hot plate, at 300°C for 90-120 minutes, as described by Di Gioacchino and Fonseca [20]. The apparatus and resulting gold speckles are shown in Figure 2. The speckle pattern was found to be remarkably sensitive to the quality of the surface. The carbon contamination due to the pre-deformation EBSD scan was found to improve the quality of the speckle pattern, forming finer and better-defined islands than those elsewhere on the prepared surface. This sharp boundary is shown in Figure 2b.

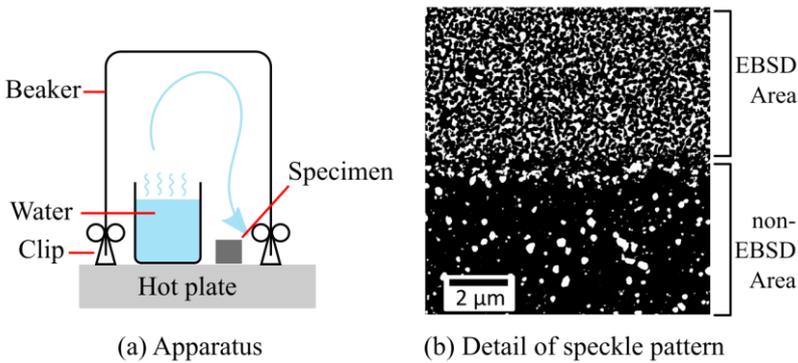

Figure 2: Diagram of the gold remodelling apparatus (a) and the resulting speckle pattern (b) (backscattered electron image), showing the effect of EBSD scanning on speckle morphology.

Backscattered electron images were captured using similar conditions to those for EBSD (20 kV, 120 μm aperture). The high atomic number of gold when compared to iron provides a high level of contrast between the steel substrate and gold speckles. Rectangular grids of images were captured with both 25% *x* and *y* overlap and stitched together using the Microsoft Image Composite Editor [39]. Again, an in-house cross-correlation DIC code in MATLAB was used to produce maps of total strain and is fully described by Jiang et al. [38].

The DIC strain maps show the effective strain, $E_{\text{eff}}$, defined as

$$E_{\text{eff}} = \sqrt{\frac{2}{3}\mathbf{E}:\mathbf{E}} \qquad (2.4.1)$$

and calculated from the in-plane components of the total finite strain tensor **E** as determined from the deformation gradient tensor **F**. Whilst no assumptions are made as to the values of the out of plane components of **E**, their values are inaccessible as out of plane displacements cannot be measured with this DIC technique.

# 3 Experimental analysis

The microstructure of the selected region is shown in Figure 3 containing several large austenite grains and three ferrite stringers. The large, central austenite grain contains three island ferrite grains as well as an annealing twin, offering an opportunity to examine any interactions between slip and ferrite whilst the large grain to the lower right allows slip in the absence of ferrite to be examined. EBSD analysis in this work and elsewhere [14] indicate no particular texture within this material. The ferrite stringers are preferentially aligned with the extrusion direction, parallel to the *x*-direction in Figure 3.

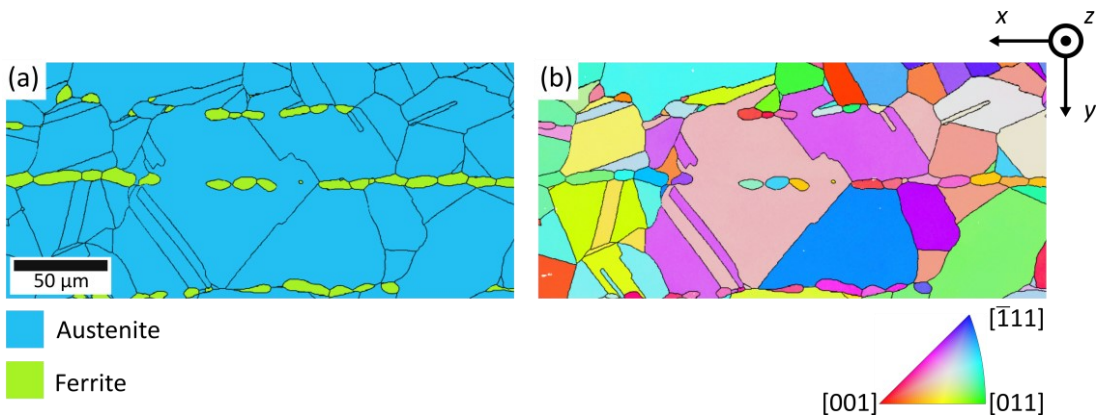

Figure 3: Microstructure of the selected region showing phase (a) and orientation (b) maps. The orientation map shows crystal directions parallel to the *z*-axis. The principal loading direction is horizontal, applying tension in the *x*-direction

Post deformation examination reveals extremely heterogenous deformation behaviour. Deformation results in high levels of GND accumulation around grain boundaries (Figure 4b), and significant residual elastic strains (Figure 4d).

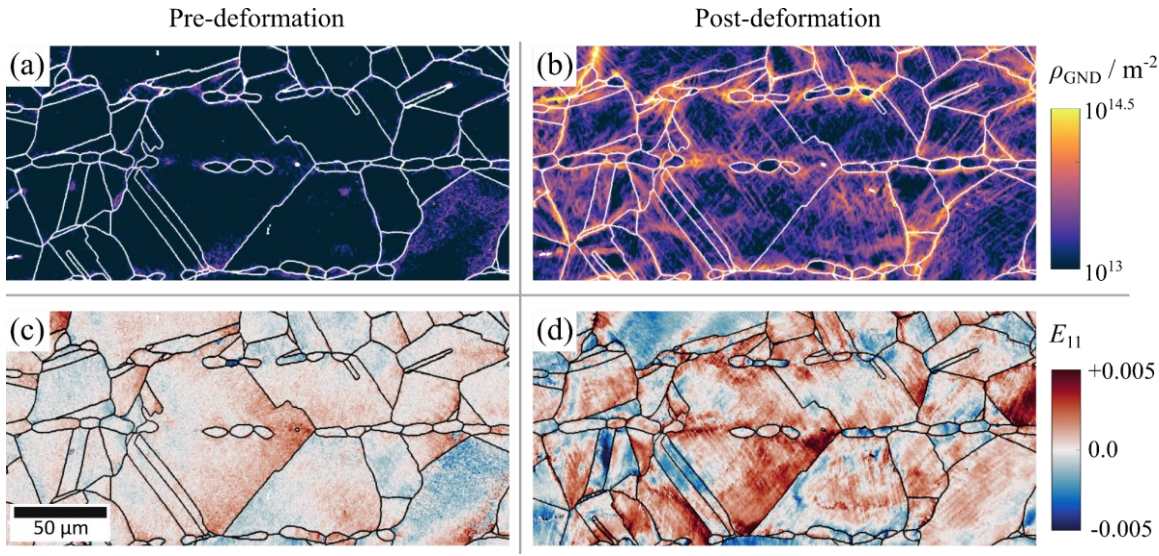

Figure 4: HR-EBSD measurements for the entire region of interest pre- (a and c) and post-deformation (b and d). Subfigures a and b show GND density, $\rho_{GND}$. Subfigures c and d show the 11 component of elastic strain, $E_{11}$ (the *x*-direction runs horizontal to the page). Elastic strain is measured in the unloaded condition with strains measured relative to a reference point within each grain.

Strain maps acquired with HR-DIC after cycle 1 and cycle 2 are shown in Figure 5. Significant slip band development is observed after the first loading cycle, with multiple slip systems being activated in the larger grains. Complex deformation patterns are observed across the region with areas displaying high levels of slip activation directly adjacent to regions showing very little deformation. Large grains show significant disparity in slip behaviour, with different regions of grains displaying different deformation behaviour simultaneously. Several interesting slip behaviours can be observed after deformation namely multiple slip system activation within single grains, a transition in slip system between cycles and the interaction of ferrite grains with slip bands. These behaviours occur at multiple points throughout the region of interest. Regions where detailed evaluation have been performed are highlighted in Figure 6.

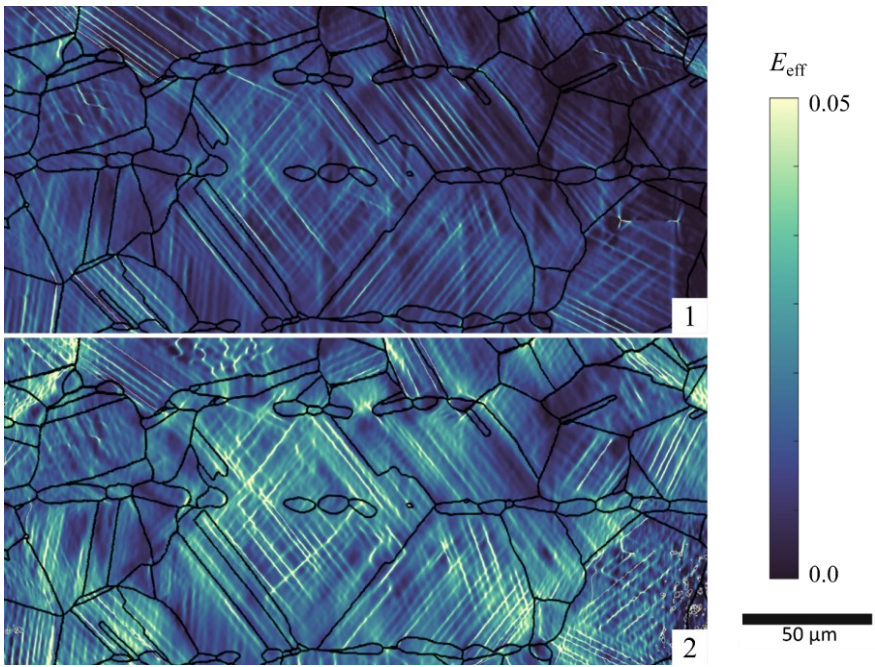

Figure 5: Effective strain maps taken over the entire region of interest, following cycles 1 and 2.

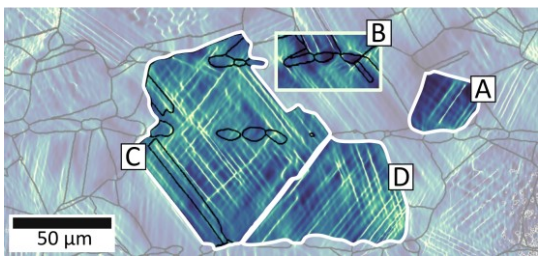

Figure 6: Regions displaying interesting slip behaviour. **A**: change in slip system between cycles. **B**: slip interaction with ferrite. **C**: Slip band interaction with island ferrite. **D**: multiple-slip system activation.

The deformation observed in Nitronic 60 is markedly different to that of 304L reported by Di Gioacchino and Fonseca [21]. Whilst the loading conditions are different to those in [21] (three-point bend here as opposed to small-scale tension) both studies conduct two interrupted deformation cycles to similar strain levels with 2% and 4% strains in the region of interest in this work and 1.6% and 6% elongation for the tensile testing in [21]. However, 304L and Nitronic 60 display significant differences in the character of deformation, specifically in the nature of slip band formation.

## 3.1 Slip band spacing and stacking fault energy

Orowan [40] predicted, after the initial onset of plasticity, slip bands in single crystals would be separated by an approximately uniform spacing of 1 μm, noting that experimentally slip bands are typically separated by 0.5 to 10 μm. Di Gioacchino and Fonseca [21] found slip bands to be separated by approximately 1 μm in 304L stainless steel. After cycle 1, slip bands form regular arrays, uniformly separated by several microns, consistent with Orowan's single crystal [40] prediction. After cycle 2, the spacing of the slip bands appears to be more irregular. Slip bands within Grain D are near uniformly spaced after cycle 1, with a separation of approximately 4 μm. After cycle 2 the average separation reduces to approximately 3 μm but the slip bands are no longer evenly spaced with the formation of tightly grouped packets of slip bands in the lower region of the grain. This behaviour is observed throughout the region of interest, with the transition from neat arrangements of slip bands after the first cycle to a less uniform arrangement after further deformation.

One key difference between the slip behaviour observed here and that observed by Di Gioacchino and Fonseca [21] is the prevalence of multiple, differently oriented slip bands within many grains. This appears to influence their separation. Another difference between this Nitronic 60 specimen and the 304L studied in [21] is the grain morphology. This specimen contains a large number of ferrite stringers as well as sharp, narrow annealing twins whereas the 304L in [21] contains little ferrite and rounded austenite grains. The microstructure has a strong effect on slip band formation, hence the formation of irregularly spaced slip band structures in this experiment in Nitronic 60.

Di Gioacchino and Fonseca [21] suggested that stacking fault energy would play a significant role in determining the deformation behaviour of stainless steel. Low stacking fault energies result in a higher prevalence of planar slip, forming wider stacking faults preventing the cross-slip of dislocations [41]. Nitronic 60 contains a significant nitrogen concentration (0.1 – 0.18 wt.% [28]) to enhance its strength and galling resistance. Nitrogen is known to reduce the stacking fault energy of austenitic stainless steels [26] but, whilst a lower SFE is considered beneficial, the exact role of stacking fault energy remains unclear in galling. Bhansali and Miller [24] proposed that a low SFE results in the formation of many stacking faults and promoting work hardening. However, Talonen and Hänninen [25] suggested that the role of stacking faults in promoting the formation of α′-martensite was more likely to be the true hardening mechanism. Ohriner et al. [2] noted that a nitrogen concentration of 0.1 wt.% is optimal for galling resistance. No α′-martensite was observed post-deformation with EBSD. Stacking faults may have formed but would not be detected if their widths were smaller than the special resolution of the EBSD measurements.

Various expressions to calculate SFE from composition exist in the literature. The expression given by Meric de Bellefon et al. [42] is used here to estimate SFE, giving the stacking fault energy in mJ m$^{-2}$ where $C_i$ are the concentrations of elements in wt.%. This relationship is valid at room temperature.

$$\text{SFE} = 2.2 + C_{\text{Ni}} - 2.9 C_{\text{Si}} + 0.77 C_{\text{Mo}} + 0.5 C_{\text{Mn}} + 40 C_{\text{C}} - 0.016 C_{\text{Cr}} - 3.6 C_{\text{N}} \qquad (3.1.1)$$

Table 1: Compositions and calculated stacking fault energies of Nitronic 60 and 304L. Compositions for Nitronic 60 were taken as the averages of the ranges given in the specification of the material supplier [43]. Stacking fault energies were calculated using equation 3.1.1.

| Alloy | Alloying component / wt.% | | | | | | | | SFE / mJ m$^{-2}$ |
|---|---|---|---|---|---|---|---|---|---|
| | Fe | C | Mn | Cr | Si | Ni | Mo | Others | |
| Nitronic 60[‡] | Bal. | 0.07 | 8.0 | 16.5 | 4.0 | 8.0 | 0.75 | 0.14 N | 5.2 |
| 304L[*] | Bal. | 0.021 | 1.96 | 18.15 | 0.34 | 9.17 | | 0.027 S, 0.031 P | 11.9 |

[‡] HP Alloys inc. [43]     [*] Di Gioacchino and Fonseca [21]

Using equation 3.1.1 with the compositions given in Table 1 gives SFEs of 5.2 and 11.9 mJ m$^{-2}$ for Nitronic 60 and 304L respectively. This reduction in stacking fault energy may lead to the change in deformation behaviour. However, both alloys would be considered low SFE alloys and the differences in deformation are likely be related to other factors. Both specimens here have similar grain sizes of the order of 50 μm. Nitronic 60 would be expected to have a higher slip strength due to the solution strengthening effect of nitrogen and slip activation would be accordingly more difficult. The Nitronic 60 specimen here appears to have a greater number of ferrite stringers than the 304L in [21] and these may play a role in slip activation and stress localisation. It is therefore likely that the local stress states generated by the presence of these ferrite stringers promotes the generation of multiple slip system activation within individual grains, outweighing the small difference in SFE between these two alloys.

## 3.2 Region A: Change in slip system between cycles

Grain A displays a complete change in active slip system between the two cycles. Single slip appears to occur during both cycles but on different slip systems. Slip bands were seen after cycle 1 and remain after the second cycle. After cycle 2, a different set of bands were observed, appearing to be unhindered by the initial slip bands (Figure 7).

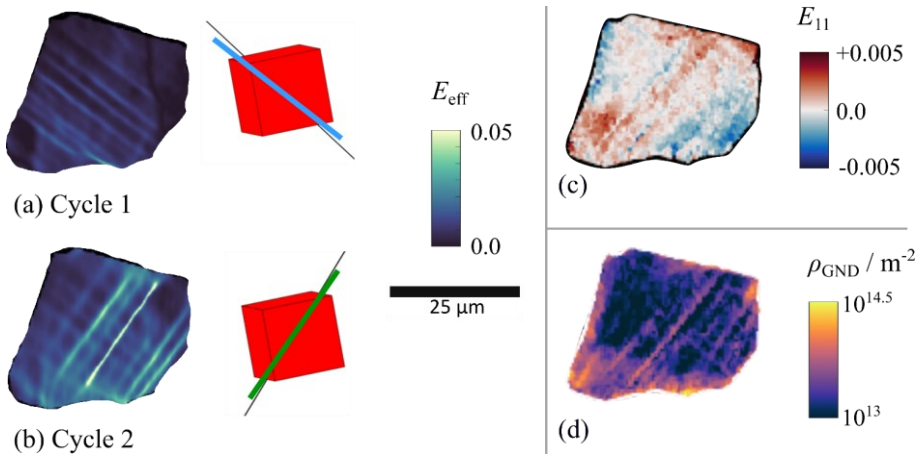

Figure 7: DIC strain maps after cycle 1 (a) and cycle 2 (b). Grain displaying change in single slip behaviour between cycles with comparison between most likely candidate for slip trace (black) and measured slip trace (colour) superimposed on a unit cell. Corresponding HR-EBSD measured elastic strain (c) and GND density (d) maps.

Global Schmid factors were a poor predictor of active slip system with values of 0.24 (6th highest, the maximum Schmid factor is 0.48) and 0.44 (2nd highest) for the possible active slip systems in cycles 1 and 2 respectively, assuming a uniaxial stress state in the horizontal ($x$) direction. This indicates that the local stress state within the region surrounding this grain is different to the global stress state and that the influence of local microstructure outweighs the externally applied remote load.

This grain is directly adjacent to a ferrite stringer (just below it) which will have a strong influence on the local stress state as ferrite has been shown in this study (and by Zhao et al. [14]) to yield less readily than austenite, providing constraint and altering the stress state within grain A. The deformation of the surrounding grains in the first cycle, and the subsequent hardening, could also change the local stress state, changing which slip system is favourable. The grains surrounding grain A (see Figure 5) show varying degrees of slip activation, likely resulting in a stress state different to the globally applied stress state.

The ferrite grains further influence the slip in the lower region of this grain in cycle 2 (see Figure 6 for position of ferrite stringer). In cycle 2, the new slip bands are prominent in the centre of the grain, bending in the region adjacent to the ferrite grain and nearly meeting the grain boundary at right angles (lower left of Figure 7b). This is similar to the behaviour observed in [21] where slip bands bend at grain boundaries due to lattice curvature. The ferrite stringer adjacent to this grain shows no significant deformation whereas the upper region of the austenite grain displays significant deformation. This mismatch could be supported by lattice curvature, hence the bending slip bands. This argument is supported by the region of elevated GND density in Figure 7d. The diffuse nature of the slip bands at the top of the grain after cycle 2 is also accompanied by elevated GND density. The blurring of the slip bands could be similar to the bending of bands in the lower region of the grain and due to lattice curvature.

## 3.3 Region B: Slip around ferrite grains

Slip bands were observed around the three ferrite grains in region B (Figure 8) accompanied by significant GND densities. The ferrite grains themselves showed little deformation with low GND densities and effective strains following deformation. Slip bands approaching austenite-ferrite phase boundaries at angles approaching 90° are blocked; this behaviour is particularly clear after cycle 1 (Figure 8a). Following cycle 2, the slip bands in the grain below the ferrite stringer were more diffuse, possibly due to the activation of a different slip system preventing the identification of discrete slip bands.

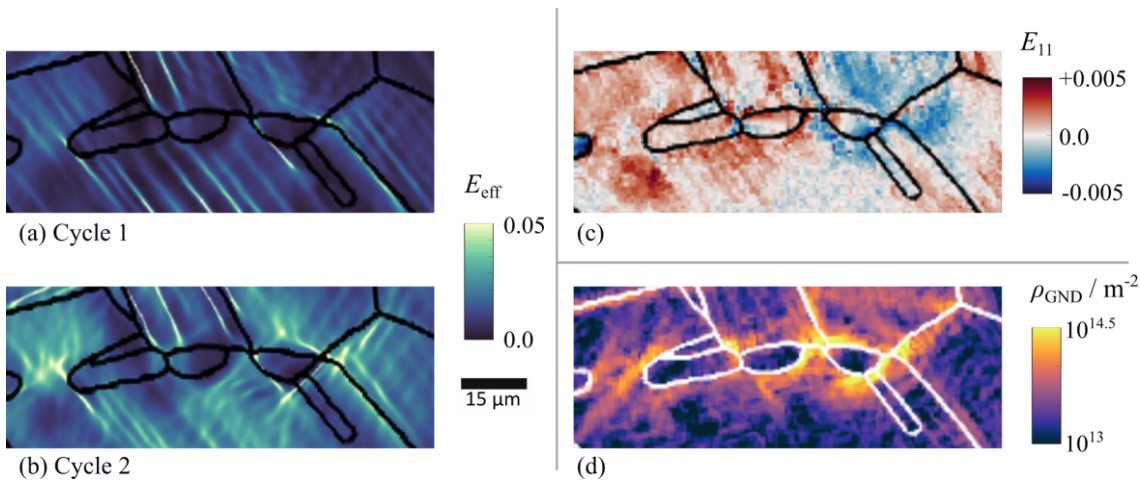

Figure 8: Slip bands surrounding ferrite grains. DIC effective strain maps are shown after cycles 1 (a) and 2 (b) with HR-EBSD measured residual elastic strain (c) and GND density (d).

The sharp slip bands in the centre of Figure 8 abruptly end at the austenite-ferrite phase boundary with the bands remaining distinct right up to the grain boundary. The GND density (Figure 8c) in this area is lower than in other areas where slip bands meet a ferrite grain. The absence of elevated GND density and the lack of any slip band bending or blurring suggests that this area has minimal lattice curvature and that there is minimal grain rotation during deformation. This may be explained by the grain morphology; this grain is an annealing twin to the grain to the right and has a long, thin shape. This constraint could prevent significant grain rotation and any significant lattice curvature at the grain boundary. The general trend of higher levels of GND densities at grain boundaries means that the presence of the small ferrite grains is of benefit in terms of hardening as this increases the total area of grain boundaries.

## 3.4 Region C: Multiple slip system activation within single grains

Region C includes an extremely large grain containing both annealing twins and three island ferrite grains as part of a longer stringer. This region demonstrates both interaction of slip systems with one another as well as slip propagation through a ferrite grain. Multiple slip systems are activated after cycle 1 with different slip systems activating in different regions of the grain (Figure 9). Parallel slip traces were observed in the twinned regions of the grain due to the orientation relationship between parent and twin grains resulting in slip planes oriented to give identical slip traces.

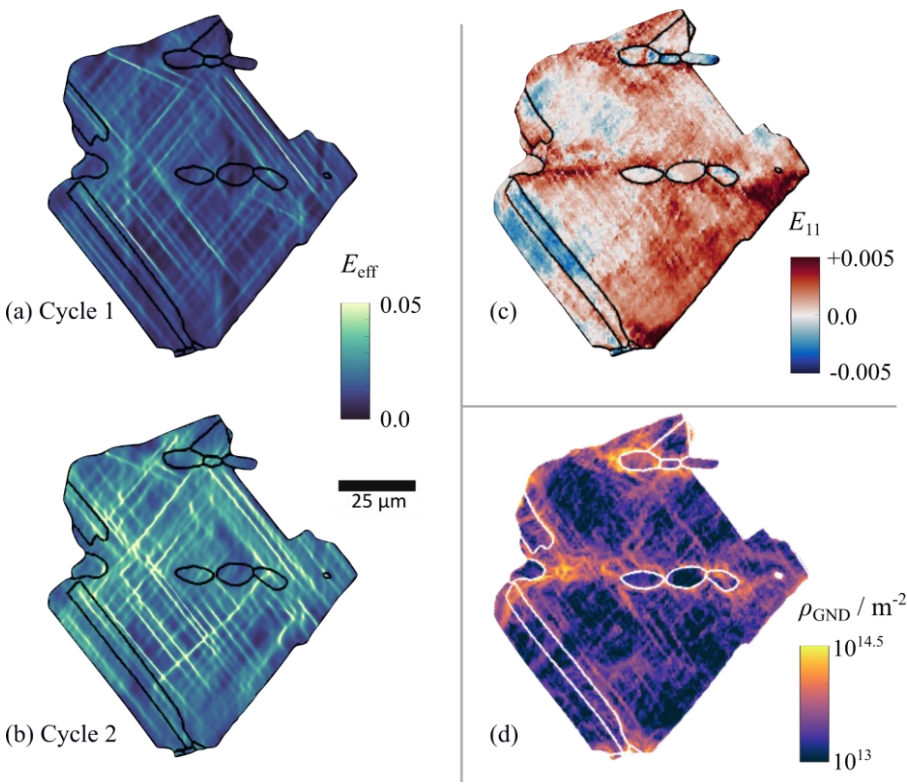

Figure 9: DIC measured effective strain maps after cycle 1 (a) and cycle 2 (b) of region C, with residual elastic strain (c) and GND density (d) maps of the corresponding region.

Slip activity increased following cycle 2; slip bands present after cycle 1 became more prominent with their strain values increasing and new slip bands developed throughout the grain. Slip bands which first developed during cycle 1 became thicker and more intense after cycle 2. Slip bands intersect at multiple points within the grain, appearing to cross each other without impeding propagation in some regions and blocking further slip completely in others. This complex slip interaction behaviour is accompanied by heterogeneous GND distributions and significant residual elastic strain gradients. This grain displays several different slip interaction behaviours, and these appear to have an influence on the resulting GND densities. The ferrite stringers contribute to the generation of GNDs and high residual elastic strains. These strains develop around the ferrite stringer, particularly between the ferrite grains and surrounding austenite grain boundary.

The upper region of this grain displays the activation of three different slip systems; possible slip systems have been identified (Figure 10a) by matching the observed slip traces to the possible traces as determined from the crystallographic orientation of this grain. Variable interaction between the three slip systems was observed, with the green and blue systems appearing to block one another whereas the blue and purple systems intersect.

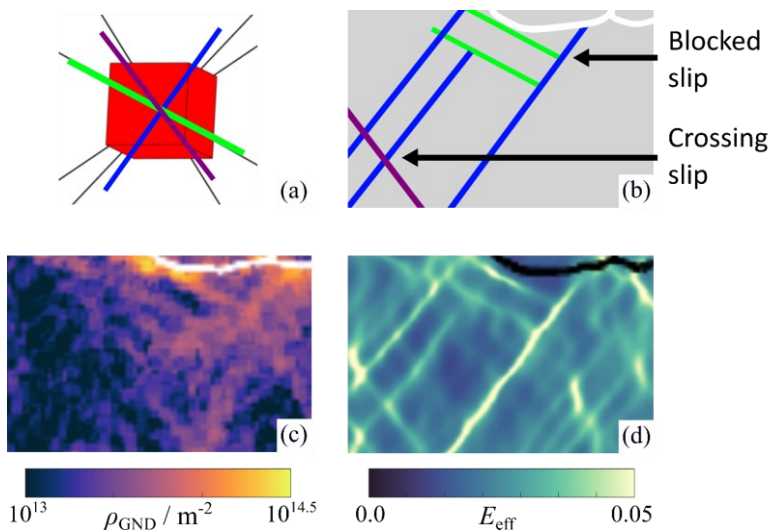

Figure 10: Region showing both crossing slip systems and slip blocking. (a) shows the theoretical slip traces in black compared with the measured traces in colour. An additional candidate slip system for the blue experimental slip trace is also shown. (b) indicates slip systems of the same type. (c) shows GND density and (d) effective strain (both after cycle 2)

Regions demonstrating slip blocking display higher GND densities than those showing simple intersection of slip bands. This suggests that the interaction of the two slip systems could result in cross hardening due to the local lattice curvature. Figure 10c shows the GND density map of the region displaying blocking and crossing slip. GND densities around areas of slip band intersection also appear to be higher than the surrounding matrix, but not as high as those associated with slip blocking behaviour. Figure 11 shows the crossing of two slip traces without blocking each other. Two strong slip bands are seen to intersect in the lower portion of this region, accompanied by increased GND density in the same region. Although there is no blocking of either slip system, the GND density within this region is higher than the region above, containing single slip.

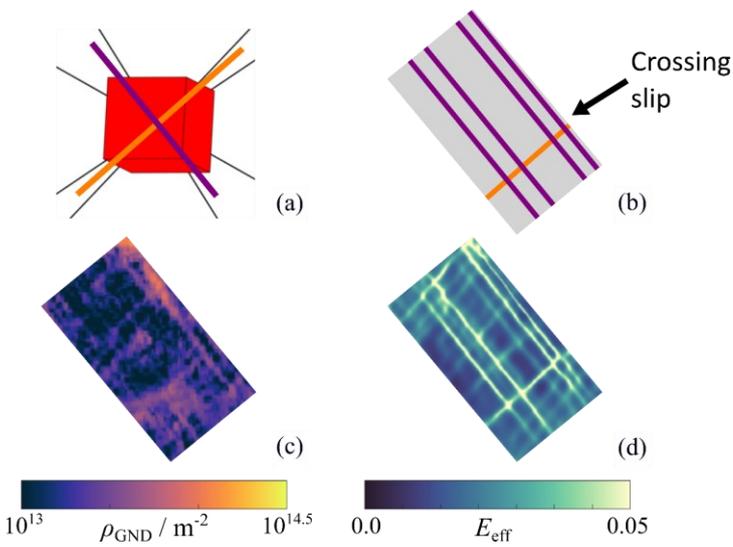

Figure 11: Region showing crossing slip systems. (a) shows the theoretical slip traces in black compared with the measured traces in colour. (b) indicates slip systems of the same type. (c) shows GND density and (d) effective strain (both after cycle 2)

Finally, this grain also demonstrates a region showing complex slip interaction behaviour (Figure 12), showing both crossing and blocked slip. The high density of interacting slip bands results in the highest density of GND within this grain, with a peak GND density of $2.3 \times 10^{14}$ m$^{-2}$. This region is also bounded by two ferrite grains. This additional constraint could lead to increased levels of strain localisation and lattice curvature, hence the accumulation of GNDs.

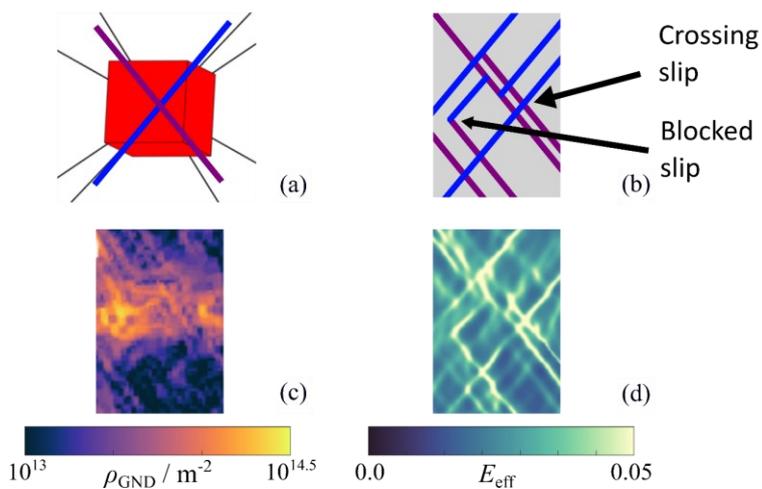

Figure 12: Region showing a complex series of slip junctions, displaying both crossing and blocking behaviour. (a) shows the theoretical slip traces in black compared with the measured traces in colour. (b) indicates slip systems of the same type. (c) shows GND density and (d) effective strain (both after cycle 2)

The presence of the island ferrite grains plays an important role in slip activation within this grain. Ferrite grains 1 and 2 (as marked in Figure 13) display slip bands passing through the ferrite-austenite grain boundaries. Grain 1 shows a single, faint slip trace whereas two strong slip traces pass from the austenite into ferrite grain 2, intersecting within the grain. The highest GND densities are coincident with the intersection of these slip bands ($\rho_{GND} = 1.4 \times 10^{14}$ m$^{-2}$ at the centre of the grain). It is difficult to ascertain whether this is a true case of slip band blocking; the proximity to the grain boundary could ultimately be responsible for the slip blocking behaviour. Nevertheless, this behaviour is accompanied by a significant accumulation of GNDs.

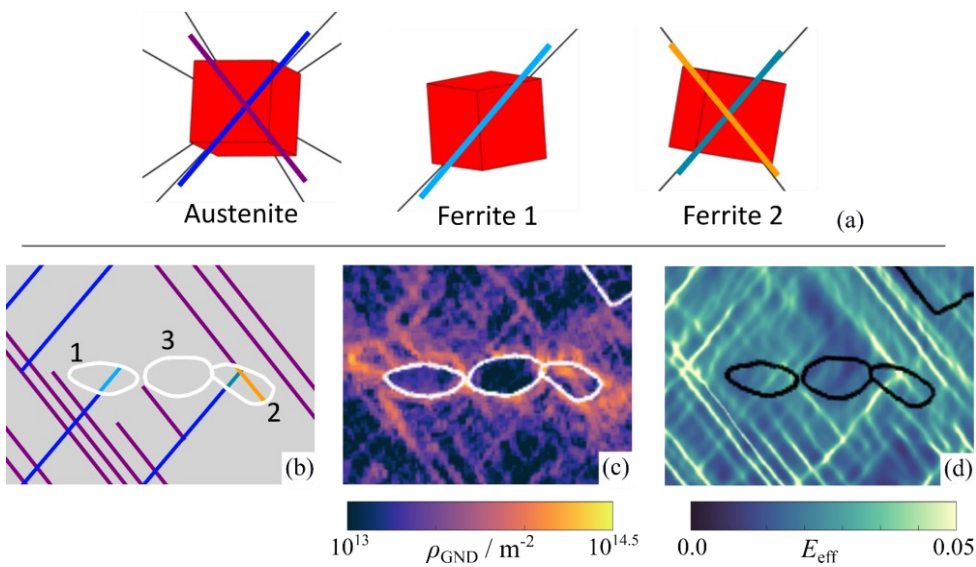

Figure 13: Slip interaction with ferrite stringer. (a) shows the orientations and predicted slip traces (black) for the austenite and two ferrite grains with measured traces in colour. (b) shows the observed traces labelled by slip system. (c) and (d) show the GND density and total strain respectively for the corresponding area (both after cycle 2).

The central ferrite grain (3) displays very little deformation, with no slip traces visible and significantly lower GND densities than those developed within the surrounding austenite or neighbouring ferrite grains. It is not possible to estimate a candidate slip system due to the large number of possible slip systems in the bcc system. One possible explanation for the variable levels of the plasticity between the three ferrite grains is that their deformation is controlled by slip transfer from the surrounding austenite grain. Slip bands in the austenite grain directly impinge on ferrite grains 1 and 2; slip transfer from the austenite to the ferrite could drive plasticity within these ferrite grains. Austenite slip bands do not visibly interact with ferrite grain 3, hence the lack of plasticity.

## 3.5 Region D: Interacting slip systems without ferrite

Region D displays interacting slip systems but without the influence of ferrite stringers. Single slip dominates during cycle 1 (SS1 in Figure 14a) with a second slip system (SS2) activating in the upper region of the grain. After the second cycle, the pre-existing slip bands thicken and increase in intensity. However, slip system 2 dominates with a substantial increase in the number and intensity of this type of slip band, generating an array of highly interlaced slip bands.

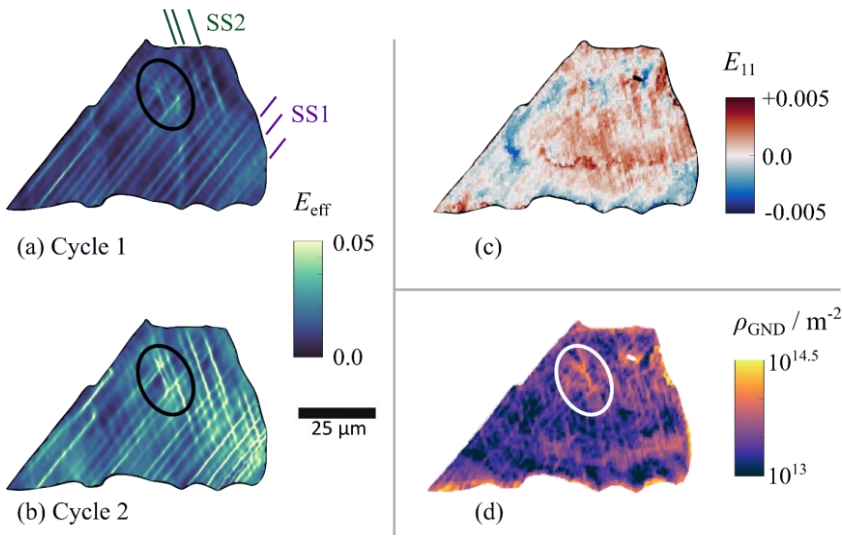

Figure 14: Multiple slip systems active within a single grain with DIC strain measurements (a) and (b) with HR-EBSD measured residual elastic strain (c) and GND density (d), both after cycle 2. The two active slip systems are labelled in (a). The circled region highlights an area displaying significant slip interaction.

The upper region of the grain contains a mixture of slip blocking and slip intersection, accompanied by significant GND density of the same magnitude as that typically found at grain boundaries. The slip interaction in the circled region appears to be more significant than that in the lower right-hand region of the grain where slip bands intersect without resulting in slip blocking. The slip interaction in the remainder of the grain does not result in the same degree of GND accumulation. A band of elevated GND density can be seen across the lower region of the grain but this is not associated with any slip band interaction.

It is unclear as to why both slip blocking and slip crossing behaviour is observed for the same two slip systems. Both slip systems are active during both cycles, as judged by either the development of new slip traces or the increase in intensity of pre-existing traces; the apparent intersection is not one set of slip traces forming over a different set of slip traces formed in the previous cycle by a now in-active slip system. The subsurface of the specimen may influence this behaviour with a hidden grain boundary or ferrite stringer just below the surface possibly responsible for these differences.

## 4 Crystal plasticity analysis

To assess the extent of GND accumulation as a direct result of slip band interaction events, a representative crystal plasticity finite element model was constructed to provide comparison with experimental results. These comparisons aimed to differentiate between the effects associated with the action of discrete slip and the effects of continuum plasticity.

### 4.1 Geometric representation and boundary conditions

The region of interest examined in the experiment was explicitly reproduced within ABAQUS, capturing surface grain geometries and extruding them through the model thickness resulting in prismatic grains. The model geometry is shown in Figure 15a. The model was meshed with C3D20R (20 noded, hexahedral reduced integration) elements with an approximate element side length of 0.5 µm to give sufficient elements per grain. The average orientation and phase from each grain as measured with EBSD (Figure 3) was specified for corresponding grains in the model.

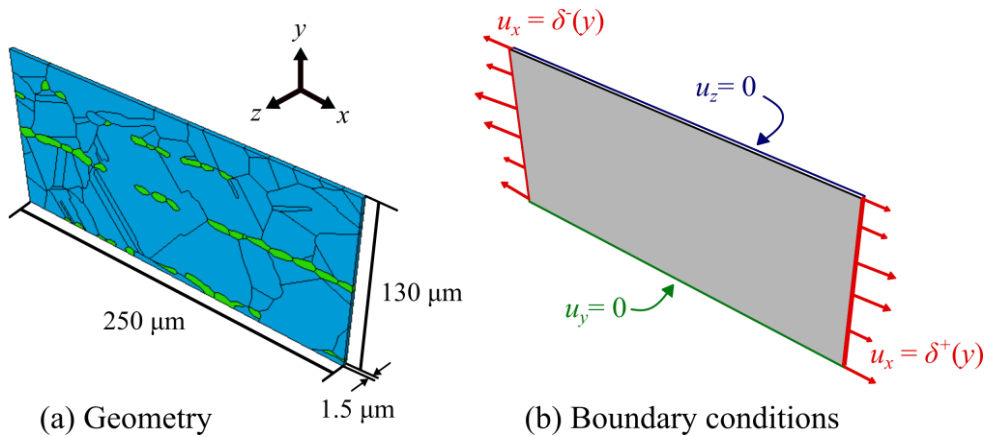

(a) Geometry       (b) Boundary conditions

Figure 15: Model geometry with dimensions (a) and boundary conditions (b). Phase colouring in (a) corresponds to that in Figure 3, with austenite blue and ferrite green. The green surface in (b) is the negative $y$-surface and the blue surface is the negative $z$-surface. Spatially varying displacement boundary conditions are applied on the red surfaces as functions of $y$-coordinate.

Plane stress boundary conditions (Figure 15b) were applied as these are representative of the free-surface condition experienced by the ROI in experiments; the front surface (positive *z*-surface) was unconstrained whilst the rear surface had motion in the *z*-direction suppressed. Motion in the *y*-direction on the bottom surface (negative *y*-surface) prevented rigid body motion with the opposite surface free in the *y*-direction. Displacement control was applied to the *x*-surfaces with displacements in the *x*-direction calculated from the DIC maps (Figure 7) using a DISP user subroutine. Displacement values on the left and right surfaces, denoted by $\delta^-$ and $\delta^+$ respectively, calculated as a functions of nodal *y*-coordinates.

The experimental DIC maps give valves of total strain and were recorded in the unloaded condition. Since the permanent plastic strains are much larger than the elastic strains, and it is the latter which are neglected with the chosen boundary condition, this is an acceptable simplification for the purposes of qualitative and mechanistic inferences of deformation behaviour.

## 4.2 Flow rule and material model

A length-scale dependent, physically-based flow rule was applied to calculate slip rates on given slip systems as a function of the local stress state [44]. The particulars of this framework are described elsewhere [44] and the salient features are described here. The plastic velocity gradient tensor $\mathbf{L}^\mathrm{p}$ is described as the sum of the slip contributions from the individual slip systems $\kappa$, with slip direction $\boldsymbol{s}^\kappa$, plane normal $\boldsymbol{n}^\kappa$, Burger vector $\boldsymbol{b}^\kappa$ and critical resolved shear stress $\tau_c^\kappa$.

$$\mathbf{L}^\mathrm{p} = \sum_\kappa \rho_\mathrm{m}\, f\, |\boldsymbol{b}^\kappa|^2 \exp\left(-\frac{\Delta F}{kT}\right) \sinh\left(\frac{(|\tau^\kappa| - \tau_c^\kappa)\Delta V}{kT}\right) \boldsymbol{s}^\kappa \otimes \boldsymbol{n}^\kappa \qquad (4.2.1)$$

The mechanism of dislocation pinning and thermally activated escape underpins the slip rule, where $\rho_\mathrm{m}$ is the density of mobile dislocation density, *f* the frequency of dislocation escape attempts, $\Delta F$ the activation energy and $\Delta V$ the activation volume. The absolute temperature is denoted by *T* and *k* is Boltzmann's constant. Slip in the austenite phase is restricted to the twelve <110>{111} type systems. In the ferrite phase, slip can occur on the twelve <111>{110} or twelve <111>{112} systems.

Hardening develops through the generation of both statistically stored and geometrically necessary dislocations. Statistically stored dislocations (SSDs) evolve during plastic deformation and result in zero-net Burgers vector, unlike geometrically stored dislocations which develop through plastic strain gradients. The critical resolved shear stress of each slip system is calculated as a function of total dislocation density, where $\tau_{c0}^{\kappa}$ is the initial critical resolved shear stress prior to deformation.

$$\tau_c^{\kappa} = \tau_{c0}^{\kappa} + |\boldsymbol{b}^{\kappa}|G\sqrt{\rho_{\text{SSD}} + \rho_{\text{GND}}} \qquad (4.2.2)$$

The SSD density evolves incrementally with time, such that

$$\dot{\rho}_{\text{SSD}} = \lambda \dot{p} \qquad (4.2.3)$$

where $\lambda$ is a hardening coefficient and $\dot{p}$ related the plastic deformation rate tensor such that $\dot{p} = \sqrt{\frac{2}{3}\mathbf{D}^{\text{p}}:\mathbf{D}^{\text{p}}}$.

The GND density determined from Nye's dislocation tensor, which is related to local plastic strain gradients [45].

$$\boldsymbol{\Lambda} = \text{curl}(\mathbf{F}^{\text{p}}) \qquad (4.2.4)$$

Due to the non-uniqueness of the relationship between lattice curvature and GND density, an $L_2$ minimisation scheme is used to calculate $\rho_{\text{GND}}$ [46]. The flow rule and associated hardening law were implemented within the commercial finite element solver ABAQUS with a UMAT user subroutine [47].

## 4.3 Material properties

Anisotropic stiffness properties and slip rule properties representative of the deformation observed experimentally are given in Table 2. The austenite phase has stiffer elastic moduli but the ferrite phase a higher critical resolved shear stress. The slip rule (equation 4.2.1) requires several physically based parameters relating the pinning and thermally activated escape of dislocations to the rate of crystallographic slip. These parameters individually affect different aspects of the rate of crystallographic slip, controlling the onset of plasticity, the rate of hardening, and any levels of strain rate sensitivity. The suitability of these parameters in capturing the behaviour both the bulk and spatially resolved strain behaviour of Nitronic 60 is demonstrated in the following sections.

Table 2: Elastic constants (Young's modulus $E$, shear modulus $G$ and thermal expansivity) and slip rule parameters (equation 4.2.1) and hardening coefficients (equation 4.2.3) for the austenite (aus) and ferrite (fer) phases.

| | Elastic constants | | | Slip rule | | | | | | |
|---|---|---|---|---|---|---|---|---|---|---|
| Phase | $E$ / GPa | $G$ / GPa | $\alpha$ / - | $|\boldsymbol{b}|$ / Å | $\nu$ / s$^{-1}$ | $\rho_{\text{m}}$ / m$^{-2}$ | $\Delta F$ / J | $\Delta V$ / m$^3$ | $\tau_c$ / MPa | $\lambda$ / μm$^{-2}$ |
| Aus | 280 | 110 | 1.3×10$^{-5}$ | 2.54 | 10$^{11}$ | 0.01 | 2.6× 10$^{-20}$ | 40$|\boldsymbol{b}|^3$ | 175 | 80 |
| Fer | 235 | 45 | 1.3×10$^{-5}$ | 2.48 | 10$^{11}$ | 0.01 | 2.6× 10$^{-20}$ | 40$|\boldsymbol{b}|^3$ | 180 | 60 |

A representative volume element comprising 125 cubic grains was used to verify that these properties correctly capture the macroscopic stress-strain behaviour of Nitronic 60. A 5 × 5 × 5 cubic arrangement of 50 μm grains was used to represent the Nitronic 60 microstructure. Ferrite grains were neglected due to the small ferrite volume fraction. A random texture (uniform sampling over SO(3) rotation space) was applied. A tensile boundary condition was applied with a strain rate of $10^{-3}$ s$^{-1}$. The model geometry and boundary conditions are shown in Figure 16a. The RVE stress strain-curve (Figure 16b) showed excellent agreement with experimental tensile testing data [14], confirming that these properties are fully representative of the behaviour of Nitronic 60.

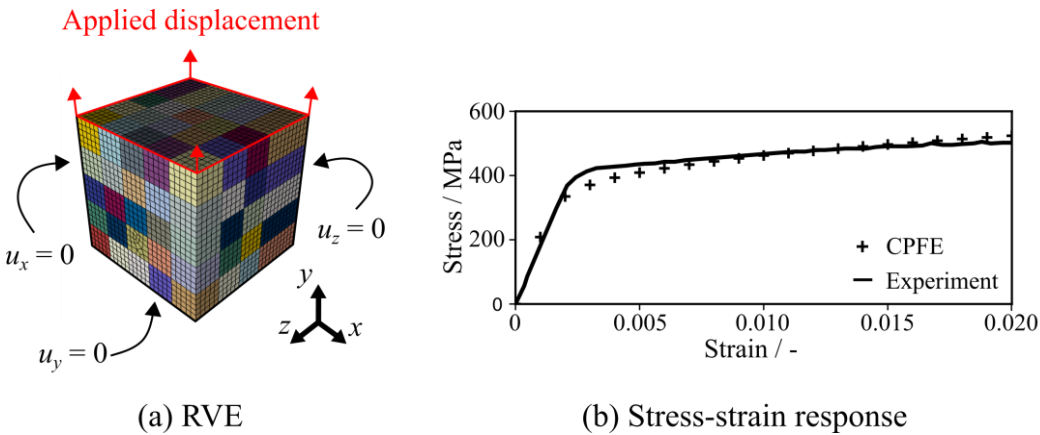

(a) RVE  (b) Stress-strain response

Figure 16: Representative volume element (a) with comparison between the resulting CPFE and experimental (after [14]) stress strain curves (b). The colours in (a) denote the different grains.

## 4.4 Results

The simulation results capture the main deformation trends observed experimentally. Slip activation is widespread throughout the model as demonstrated in Figure 17a. The slip activations and deformation heterogeneity are reasonably well captured with grains displaying lower levels of plasticity experimentally (Figure 17b) than for the corresponding values in the numerical results. Ferrite grains in the simulation display varying levels of plasticity between individual ferrite grains. As with experiment, ferrite grains neighbouring highly deformed austenite grains display higher levels of plasticity. Simulation results show statistical correspondence with experimentally measured strain values. Histograms of the *xx*-strain component (Figure 18) show agreement between experimental and simulation strain values. DIC measured strain values show a slight shift to higher values and narrowing of the curve. This is related to the presence of discrete slip bands, manifested as narrow regions of elevated strain. Their absence in the continuum CP model is responsible for this shift.

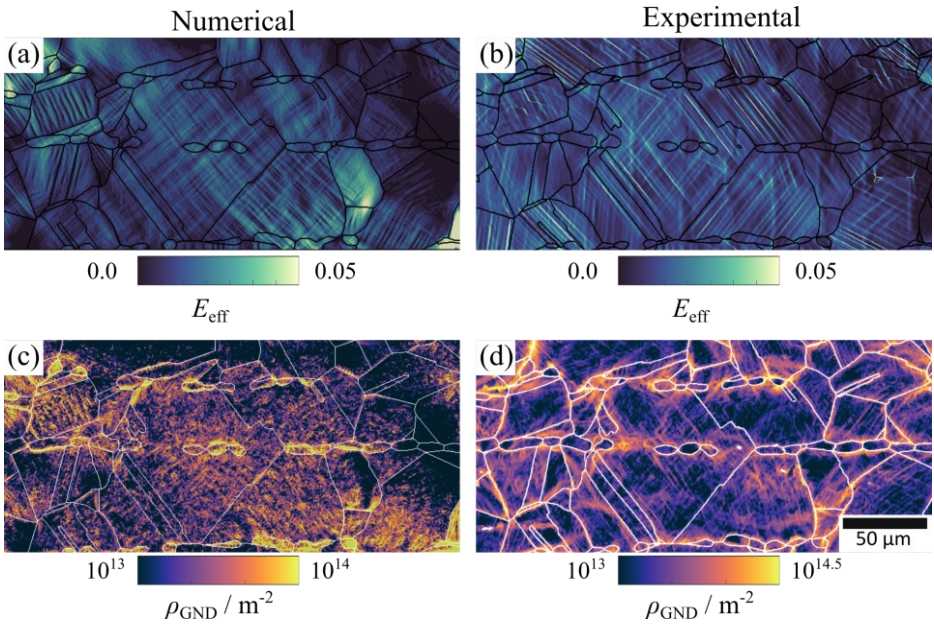

Figure 17: Comparison between the numerical and experimental results for effective strain (a and b) and GND density (c and d). Note the difference in colour scale limits in c and d.

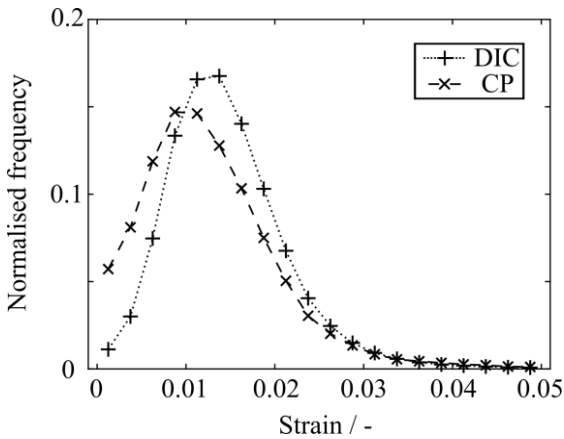

Figure 18: Histograms comparing values for the xx strain component (primary loading direction) for experimental DIC and crystal plasticity (CP) simulation results, corresponding to the strain maps Figure 17a and Figure 17b respectively.

The numerical model correctly captures similar orders of magnitudes of GND densities (Figure 17c) with concentrations around grain boundaries and ferrite grains as expected. However, the GND fields are more diffuse than observed experimentally, forming relatively homogeneous clouds of GND density in the central regions of the larger grains. Smaller grains display more heterogeneous GND distributions, with grain boundary effects dominating. Absolute values of GND densities are slightly lower in the numerical results than the experimental results, however the overall trends and distributions are well captured.

Ferrite grains displayed a higher level of GND density in the simulation than experiment. Experiments demonstrated the interaction between austenite slip bands and ferrite grains was important in determining the degree of deformation within these grains, with slip bands traversing some ferrite-austenite grain boundaries but not others (see Figure 13). This effect was not observed in the simulations since this highly localised deformation behaviour cannot be captured with CPFE.

Regions C and D (see Figure 6) displayed prominent GND fields surrounding slip band interaction areas. A detailed comparison (Figure 19) shows that regions of peak GND density are in the same locations as those displaying GND density peaks and slip interactions. Simulation results show more diffuse regions of GNDs, suggesting that the experimentally observed high concentrations are due to discrete slip band interactions. However, the model does correctly reproduce the localisations of lattice curvature in these regions.

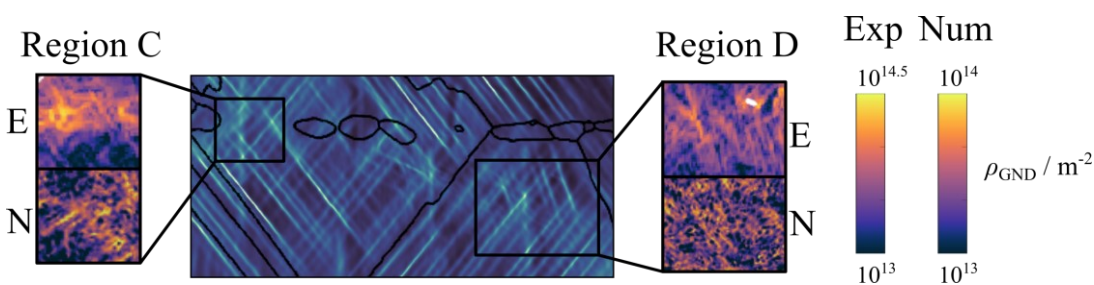

Figure 19: Detailed comparison between experimental and simulation GND density plots for regions displaying slip interactions. E and N denote experiment and numerical simulation results respectively. The central strain map shows DIC measurements displayed in Figure 17b. Again, note the different limits in colour scale

These results show that general trends in deformation are well captured by the crystal plasticity model. However, the experimental results display higher levels of GND density localisation. Since the key difference between the simulation and the model is the absence of discrete slip activity (as the CPFE model homogenises this effect) the interaction of slip bands is the probable cause of this highly localised behaviour. These slip system interaction effects have been examined in an separate study, applying the same crystal plasticity modelling framework as this work to examine the nickel-base superalloy RR1000 [48]. This investigation in RR1000 demonstrates similar behaviour with slip band interactions resulting in latent hardening and quantifies localised peaks in GND density. These independent observations show good agreement with both the experimental and simulation results described herein, showing that these behaviours are material independent and are inherent in slip-mediated plasticity.

# 5 Discussion

The deformation trends observed in Nitronic 60 are highly complex and differ from those reported for 304L [21]. The main difference is the propensity for the activation of multiple slip systems within single grains in Nitronic 60 as opposed to the largely single slip in 304L. The large amount of multiple slip results in complex slip band interactions which dominate the deformation. It is not clear as to why multiple slip system activation is common in Nitronic 60 but planar slip dominates in 304L. However, the lower stacking fault energy of Nitronic 60 may result in a higher rate of work hardening within the material requiring a secondary slip system to activate to accommodate deformation whereas the activation of one main slip system suffices in 304L. A second factor is the initiation of slip due to similar slip systems in neighbouring grains. Dislocations reaching a grain boundary may cause a sufficient stress concentration in the neighbouring grain to activate a similar slip system in the neighbouring grain in addition to any already active slip systems. Slip band-grain boundary interactions generate significant degrees of hardening, with concentrations of GNDs decorating all grain boundaries in the experimentally observed region of interest. As discussed elsewhere [14], the presence of annealing twins increases the total area of grain boundaries throughout the microstructure, increasing the level of hardening through the generation of associated GNDs. The observed slip band-slip band interactions are generally found away from grain boundaries, towards the centre of grains and this effect can therefore be considered distinct from slip band-grain boundary interaction effect. Both effects act to drive up the work hardening rate of Nitronic 60 through the generation of GNDs.

There are therefore three primary latent hardening mechanisms operative within this microstructure: slip band-slip band interactions, slip band-phase boundary interactions at the austenite-ferrite phase boundaries, and slip band-grain boundary interactions at the austenite-austenite grain boundaries. Each of these generates significant degrees of latent hardening. The presence of small, dispersed ferrite grains throughout the microstructure provides increased opportunity for local hardening with their presence largely beneficial in resisting excessive plastic deformation. Whilst found in the form of elongated stringers by virtue of the extrusion processing route, a more uniform dispersion of ferrite would provide a more homogeneous hardening effect. The large number of grain boundaries, further increased by the presence of annealing twins, increases the likelihood of slip band-grain boundary interactions and the associated hardening. Promoting an increase in the total number of grain boundaries by reducing the grain size and increasing the number of twins would further strengthen this effect. However, this is somewhat at odds with generating slip band-slip band interactions as these are more prevalent in larger grains where multiple slip systems interact. From a mechanical performance view, promoting a larger number of grain boundaries and a uniform distribution of ferrite grains would likely be the primary strengthening mechanisms based on the frequency of these events occurring, as these effects occur at many grain boundaries. Whilst less frequently encountered and found at discrete points within the microstructure, GND generation at slip band-slip band interactions is nonetheless a beneficial hardening mechanism and will occur throughout the material.

Crystal plasticity analysis has demonstrated that slip band interaction events are responsible for the localised regions of elevated GND density. In the experimental results, regions with multiple slip activation are associated with higher strain gradients, generating elevated GND densities. In the simulations, localised slip activation generated similar strain gradients, causing corresponding elevated GND densities. Whilst the simulation dislocation density fields are more diffuse, the simulation correctly captures these elevated GND densities when compared to other regions undergoing lower levels of deformation.

The different slip interactions follow a general trend of increasing GND density with increasing intensity of slip band interaction. All dislocation activity results in some degree of GND accumulation. Planar slip with a single slip system active within a region results in the smallest increase in GND density. Double slip activation within a region where the slip systems intersect with minimal interaction (i.e. the intensity of the slip bands is largely unaffected) results in moderate GND densities. Finally, the intersection of two slip systems where there is a degree of slip band blocking results in the highest GND densities.

The blocking of slip bands by other slip bands is fundamentally different to the blocking of slip bands by grain boundaries. The major barrier for the passage of dislocations (and the formation of intergranular slip bands) is the misorientation between the grains. Various studies have considered the transmission of dislocations through grain boundaries, relating slip band transmission to the relative orientation of possible slip planes. Measures have been developed to allow the propensity for slip transfer through a grain boundary to be quantified as functions of misorientation between slip planes and resolved shear stresses [49–52]. However, within a single grain, this misorientation is largely absent. For the case of slip bands blocking other slip bands, there is minimal misorientation across the boundary as these events occur within single grains; whilst misorientation within grains does develop during deformation, there is a continuous change in orientation through a grain and not a discontinuity in orientation as at a grain boundary. This clearly demarcates the herein described slip band-on-slip band interactions from slip band-grain boundary interactions.

The mechanisms of formation of these GND structures surrounding blocked slip bands could be related to the constraint provided by the blocking slip band. For planar slip, a single slip system with slip direction $s_1$ and plane normal $n_1$ is active, forming parallel slip bands. The surrounding grains provide constraint, but these are generally well removed from the slip band, leading to low levels of lattice curvature and low GND densities. Regions with single, planar slip far from grain boundaries would be expected to show the lowest increases in GND density and single slip would not cause significant degrees of lattice curvature without the constraint of neighbouring grains (in the manner of a single crystal able to freely deform).

In contrast to this behaviour, blocking slip results in the highest accumulation of GNDs. Consider the primary slip band (S1) blocking a secondary slip band (S2) with slip direction $s_2$ and normal $n_2$ (Figure 20c). The blocking slip band, S1, acts as a barrier to the S2 slip band, blocking the passage of dislocations. The region below the blocking slip band contains no active slip systems and therefore does not plastically deform, however, continuing deformation causes further slip activity in the region above band S1.

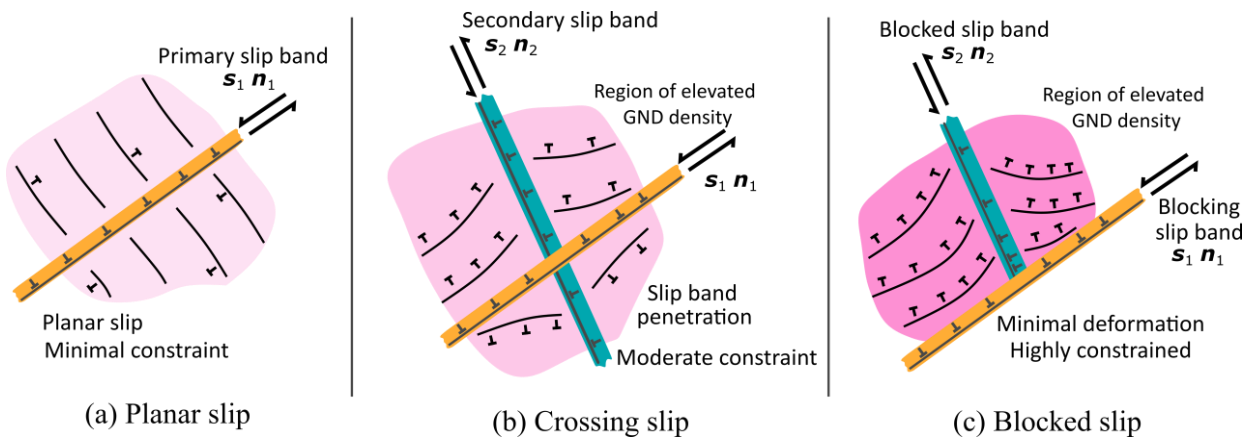

(a) Planar slip    (b) Crossing slip    (c) Blocked slip

Figure 20: Schematic representation of the generation of lattice curvature and geometrically necessary dislocations (pink regions) associated with differing levels of slip interaction. Planar slip with a single slip system described by normal $n_1$ and direction $s_1$ (a) results in minimal lattice curvature due to the lower levels of constraint. Regions where there are two slip systems active with minimal interaction between the two slip bands (b) generate moderate levels of lattice curvature. The constraint generated by the blocked slip band (c) results in the highest levels of lattice curvature.

Since the region below the blocking slip band has no slip activity, it remains undeformed and provides constraint to the region above the blocking band to maintain compatibility. The combination of further deformation and slip activity close to a non-deforming region results in significant lattice curvature and a corresponding increase in GND density.

In regions where slip bands intersect without blocking one another (Figure 20b), an intermediate GND density is found. A similar scenario can be imagined to that of the blocking slip band situation except with less constraint. Again, consider a secondary slip band impinging on a primary slip band, except now the secondary band is not blocked and passes through the primary band. Therefore, both sides of the primary slip band can deform plastically. Some degree of lattice curvature (and corresponding GND density) would be expected in this region to maintain compatibility but less than for the case of a blocked slip band due to the lesser degree of constraint.

Measurements of GND densities along paths in regions displaying slip band interactions are shown in Figure 21. These two regions highlight the general trend in increasing slip interaction resulting in increasing levels of lattice curvature and GND density. Figure 21a highlights a section within Region C, with a path described by the line A-A′ running approximately parallel to a slip band. Two small slip band interactions are observed (white arrows) at approximately 20 and 35 μm along A-A′. These correspond to modest increases in GND density above the background. The intersection of two major slip bands at 55 μm along the path is accompanied by a substantial increase in GND density in the surrounding region. The maximum GND density value in this region is $5.1 \times 10^{13}$ m$^{-2}$.

The GND density associated with blocking slip, as show in Figure 21b is further elevated than that associated with crossing slip in Figure 21a. The path B-B′ runs parallel with a prominent slip band in the grain in Region D. This slip band is blocked by series of slip bands at approximately 50 μm along the path. This is accompanied by large region of elevated GND density with a peak value of $1.6 \times 10^{14}$ m$^{-2}$, in excess of three times the maximum value for the crossing slip behaviour. The GND density shows a sudden decrease in the region after the slip blocking, suggesting that this region has lower levels of deformation than the region before the blocking band. This further strengthens the arguments relating lattice curvature and slip blocking in Figure 20c. These measurements show that there is a significant and quantifiable increase in GND density with increasing slip band interaction.

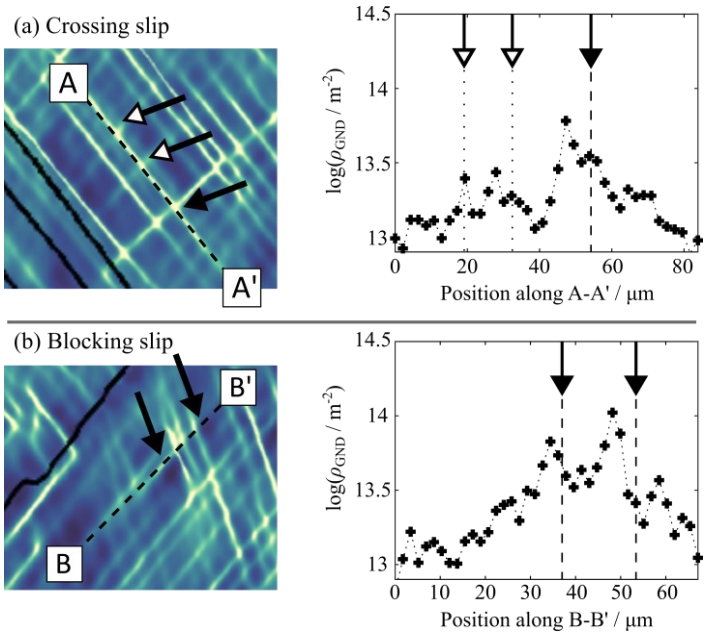

Figure 21: DIC strain maps (subsets of Figure 7 after cycle 2) to highlight slip interactions with corresponding GND density measurements along paths in regions demonstrating crossing slip (a) and blocking slip (b). In (a) the path A-A′ displays one main slip crossing event (black arrow) and two less intense slip crossing events (white arrows). In (b) the path B-B′ shows a series of strong slip interactions terminating in a slip blocking event. This region is highlighted with black arrows.

For the case of blocked slip bands, one would expect similar stress fields to those found where slip bands interact with grain boundaries, resulting in dislocation pileups. Whilst the mechanism causing the blockage of slip bands is different to the interaction of slip bands at a grain boundary, the pileup of dislocations along the slip band and the related stress fields would be expected to be qualitatively similar. The theoretical work of Eshelby et al. [53] predicted that these pileups are accompanied by a stress field to maintain mechanical equilibrium. The magnitude of the shear stress resolved onto the active slip plane is proportional to the inverse of the square root of the distance from the start of the pile-up (i.e. stress follows a $r^{-1/2}$ relationship where $r$ is the distance from the start of the pile-up). This has been experimentally verified in hexagon [16,17] and cubic [54] alloys

An attempt to calculate the strain and stress fields directly adjacent to blocked slip bands observed in this work was made, using the HR-EBSD elastic strain measurements. However, it was difficult to extract any useful information due to both the spatial resolution at which the HR-EBSD measurements were taken and the more complex slip behaviour than the simple slip band to grain boundary behaviour observed in the studies in titanium [16,17] or nickel [54]. The spatial resolution of the HR-EBSD in this work was optimised to cover a large area in order to capture a range of slip behaviour, using a step size of 0.5 μm, larger than the 0.2 μm used in [16,17] but in keeping with the work of others [14,31,55]. The activation of multiple slip systems and close proximity of several slip bands adds significant complexity when compared to the blocking of a slip band by a grain boundary, making inferences from the strain fields surrounding slip interaction events difficult. However, the general trends in the elastic strain maps (Figure 4c and d for example) show that slip band interactions do contribute the highly heterogeneous residual elastic strain fields.

The highest GND densities were found in regions with several occurrences of slip band interaction in close proximity to one another; in this study, the region detailed in Figure 12 displayed the highest GND density of $2.3 \times 10^{14}$ m$^{-2}$. The additional constraint due to multiple slip bands interacting exacerbates the effects detailed in Figure 20 forming multiple regions undergoing deformation directly adjacent to regions with little deformation. This gives rise to extremely high levels of localised GND accumulation, causing high levels of local material hardening.

Evidence of significant lattice curvature was also observed in the bending of slip bands at the edges of some grains, accompanied by a region of elevated GND density. Di Gioacchino and Fonseca [21] propose two mechanisms for this behaviour. In the first mechanism, GND formation supports the lattice curvature required to maintain compatibility with a neighbouring grain resulting in bent or diffuse slip bands. The second mechanism relies on the activation of a secondary slip system near the grain boundary to resolve the incompatibility, resulting in a separate domain of deformation near the grain boundary. The observation of curved or diffuse slip bands here suggests that the first mechanism is more likely.

The absence of strain induced martensite formation is interesting in the context of galling. One of the key purported mechanisms for the galling resistance of iron-base hard facings is the formation of α′-martensite but this has not been observed in this study. It would be expected that this transformation would be likely at room temperature, as this change is suggested to cause the well documented poor galling resistance in stainless steels at elevated temperatures [6,10–12]. The strains observed here may not be sufficient to induce the $\gamma \rightarrow \alpha'$-martensitic transformation; strains encountered during sliding wear under galling conditions would be far greater than those experienced under three-point bend testing.

The ferrite stringers present in this Nitronic 60 specimen play a key role in deformation. Their primary effect is to provide additional grain boundaries, promoting the generation of GNDs which are known to form at the interfaces between grains to maintain compatibility. However, island ferrite grains result in the formation of GNDs away from grain boundaries in the centre of grains. Ferrite grains have been shown to be variable in terms of deformation, with some grains displaying clear slip bands directly adjacent to ferrite grains remaining completely undeformed. In general, these ferrite grains appear to act as barriers to deformation, promoting the accumulation of GNDs and providing a hardening mechanism. Ferrite stringers are pervasive throughout the microstructure. Although the ferrite volume fraction is only approximately 3% (as measured on the free surface using EBSD [14]), they appear to be reasonably uniformly dispersed throughout the entire microstructure, with a separation of around 50 μm or (typically) two to three austenite grains. As such, the ferrite stringers are important to the overall hardening behaviour due to their presence throughout the microstructure of this alloy.

The high number of ferrite stringers makes it difficult to truly disentangle the effect of the ferrite grains on deformation and the deformation inherent to the large austenite grains. Whilst large grains not containing ferrite stringers have been examined here, subsurface ferrite may influence the behaviour observed. This behaviour has also been reported in 304L where subsurface ferrite stringers caused surface slip bands to fade and become diffuse [21].

In terms of galling resistance, the open microstructure of Nitronic 60 is far from ideal. Whilst the cross-hardening due to interacting slip systems and presence of ferrite stringers both cause significant GND evolution and work hardening, a finer microstructure would provide a greater resistance to gross plastic deformation. To resist galling, a material must possess a sufficiently high work hardening rate to resist the high levels of localised deformation associated with sliding contact. A finer microstructure with an even dispersion would provide a significant degree of hardening rather than the large austenite grains and smaller ferrite grains found here. The microstructure of the Nitronic 60 considered here is due to the casting process used to produce this specimen. Using a powder metallurgy process to produce a Nitronic 60 specimen may produce a material with superior properties and resistance to plastic deformation.

# 6   Conclusions

The deformation of the galling resistant stainless steel Nitronic 60 has been studied with correlated HR-EBSD and HR-DIC, providing detailed insight into the complex slip interaction behaviour as well as quantitative measures of strain and geometrically necessary dislocation densities. The key conclusions are as follows.

(1) The deformation of this material is shown to be extremely complex and non-homogeneous. Multiple slip systems are found to be active within single grains, even at low levels of strain, resulting in the formation of multiple, differently oriented slip bands

(2) Ferrite stringers play a key role in hardening, causing significant evolution of geometrically necessary dislocation density.

(3) Slip band interactions have been demonstrated to be responsible for highly localised regions of GND density, causing significant local hardening.

(4) Differing degrees of slip band interaction result in different degrees of lattice curvature and GND. Planar slip bands result in low levels of GND density. Crossing slip bands result in intermediate GND densities. Blocking slip bands are associated with the highest GND densities. This is rationalised by considering the varying levels of constraint and the corresponding lattice curvature for the three cases.

This study shows the deformation of even simple stainless steels is highly heterogeneous in nature and these heterogeneities need to be considered when designing alloys for industrial purposes.

# 7 Acknowledgements

The authors gratefully acknowledge the support of Rolls-Royce plc., the Engineering and Physical Sciences Research Council (EPSRC) and the ICO Centre for Doctoral Training in Nuclear Energy (EP/L015900/1). The authors would also wish to thank Mr Ben Wood for machining the modified three-point bend grips, Dr Ben Britton for providing the HR-EBSD and HR-DIC MATLAB codes and Dr David Stewart for useful discussions and insight into hard-facing alloys. All electron microscopy was performed at the Harvey Flower Electron Microscopy Suite at Imperial College London.

# 8 Data Availability

The raw/processed data required to reproduce these findings cannot be shared at this time due to technical or time limitations.